\def\fun#1#2{\lower3.6pt\vbox{\baselineskip0pt\lineskip.9pt
  \ialign{$\mathsurround=0pt#1\hfil##\hfil$\crcr#2\crcr\sim\crcr}}}
\relax
\documentstyle[12pt]{article}
\advance\textheight by \topskip
\begin{document}
\rightline{hepth@xxx/yymmnn}
\vskip 1cm
\begin{center}
{\Large\bf  AXION-DILATON BLACK HOLES \footnote{Talk presented at
the
TEXAS/PASCOS conference, Berkeley, December 1992}}\\ \vskip 0.9 cm
{\bf Renata Kallosh \footnote { \  E-mail:
kallosh@physics.stanford.edu}}
 \vskip 0.5cm
Physics Department, Stanford University, Stanford   CA 94305
\end{center}
\vskip .5 cm
\centerline{\bf ABSTRACT}
\begin{quotation}
In this talk  some essential features of stringy black holes are
described.
We consider charged four-dimensional
axion-dilaton black holes.
The Hawking temperature and the entropy of all solutions   are shown
to be
simple functions of the squares of supercharges, defining the
positivity
bounds.  Spherically symmetric and multi black
hole solutions are presented. The extreme solutions have some
unbroken
 supersymmetries.  Axion-dilaton black holes with  zero entropy and
zero area of the horizon form a family of stable particle-like
objects,
which we call {\it holons}. We discuss the possibility of  splitting
of
nearly extreme black holes into holons.

\end{quotation}
\vskip 1 cm
This talk is based on various results obtained in collaboration with
A. Linde, T. Ort\'{\i}n, A. Peet, and
        A. Van Proeyen \cite{US} - \cite{FUTURE}
We have found general  form of charged $U(1)$  and  $U(1)\times U(1)$
four-dimensional static  axion-dilaton black hole
solutions. In particular our solutions include dilaton black holes
\cite{G}
and dual dyon black holes \cite{STW}.
The  independent parameters characterizing our  solutions
 are the black hole mass, electric and
 magnetic charges and  the values
of dilaton and axion fields at infinity. The dilaton and axion
charges
of our solutions are functions of these independent parameters.

Our black hole solutions have an important property which is the
consequence of
$SL(2,R)$ symmetry equations of motion following from the low-energy
string
effective action. Under $SL(2,R)$ electric-magnetic
duality rotations only the ``hair" (charges
and asymptotic values of the fields) of our solutions is
transformed.
The functional form of the solutions is duality-invariant. Dirac
quantization
of electric and magnetic charges breaks this symmetry down to
$SL(2,Z)$
\cite{KO}.

Extreme solutions have some unbroken supersymmetry, when embedded in
the
theory with local $N=4$ supersymmetry. Some solutions have unbroken
$N=1$ and some have unbroken $N=2$ supersymmetry. Solutions with
unbroken $N=1$ supersymmetry saturate one supersymmetric Bogomolny
bound
and those with unbroken $N=2$ supersymmetry saturate both bounds.
The
thermodynamical properties of stringy black holes are described by
Hawking temperature and entropy of the black holes.
The temperature and the entropy of a whole family of
axion-dilaton black holes have a remarkably simple
expression in terms of SUSY bounds \cite{FUTURE}:
\begin{eqnarray}
T &= & {1 \over 2\pi M} \;{ {\cal B}_1 \; {\cal B}_2 \over \left
({\cal B}_1 +
 {\cal B}_2 \right )^2} \ ,\nonumber\\
\nonumber\\
S &=& \pi  \left ({\cal B}_1 + {\cal B}_2 \right )^2\ ,
\label{TS}\end{eqnarray}
where two SUSY bounds for charged axion-dilaton black holes are given
by
\begin{eqnarray}
{\cal B}_1 &=&  \sqrt {M^2 - |z_1|^2}\ , \nonumber\\
\nonumber\\
{\cal B}_2 &=&  \sqrt {M^2 - |z_2|^2} \  .
\label{B}\end{eqnarray}

In eq. (\ref{B}) $M$ is the mass of the black hole and
$z_1, z_2$ are two complex
combination of electric and magnetic charges. In the global SUSY
algebra
associated with asymptotically flat space of our black holes these
combinations
of charges play the role of central charges.
The parameter $r_0$, which vanishes when the black hole becomes
extremal, is given by the product of supersymmetry bounds:

\begin{equation}
r_{0}^2 =  \frac{{\cal B}_1^2 \;{\cal B}_2^2}{M^2}=\frac{(M^2 -
|z_1|^2)\,
(M^2 - |z_2|^2)}{M^2} \  .
 \end{equation}

Consider few examples of eqs. (\ref{TS}). Schwarzschild  solution
corresponds
to $z_1 = z_2 =0$.  Thus we get from (\ref{TS})
\begin{eqnarray}
T_{Sch} &= & {1 \over 8\pi M} \nonumber\\
\nonumber\\
S_{Sch} &=& 4 \pi M^2\ ,
\end{eqnarray}
The Reissner-Nordstr\"om solution is given by $z_1=0, z_2 = q$.
Substitution
of these values of central charges into eqs. (\ref{TS}) gives
\begin{eqnarray}
T_{RN} &= & {1 \over 2\pi } \;{ \sqrt {M^2 - q^2}  \over \left (M +
 \sqrt {M^2 - q^2} \right )^2}\nonumber\\
\nonumber\\
S_{RN} &=& \pi  \left (M +
 \sqrt {M^2 - q^2} \right )^2\ .
\end{eqnarray}
Extreme Reissner-Nordstr\"om black holes with $M=q$ have zero
temperature
and  non-zero entropy, $N=1$ supersymmetry being unbroken \cite{GH}.

The charged  $U(1)$  axion-dilaton black
holes \cite{G} -\cite{STW} have $|z_1| = |z_2| = |q/ \sqrt 2|$ and
therefore
only one SUSY bound ${\cal
B}_1={\cal B}_2$. The corresponding expressions for the temperature
and the
entropy are
\begin{eqnarray}
T_{U(1)} &= & {1 \over 2\pi M} \;{ M^2 - |q / \sqrt 2|^2 \over 4 (M^2
-
|q/ \sqrt 2|^2
)}= {1 \over 8\pi M} \ ,\nonumber\\
\nonumber\\
S_{U(1)} &=& 4 \pi  \left (M^2 - |q / \sqrt 2|^2 \right )^2\ .
\end{eqnarray}

In the extreme limit $M = |q / \sqrt 2|$ the temperature remains
equal to the
temperature of the  Schwarzschild black hole and the entropy
vanishes.

Finally consider an $U(1)\times U(1)$ four-dimensional
axion-dilaton black holes, with one vector and one axial vector
field.
The moduli of central charges in terms of
electromagnetic charges are given by
\begin{eqnarray}\label{central}
|z_1|^2 &=& \frac{1 }{ 2}(P_1^2 + Q_1^2 + P_2^2 + Q_2^2) - E  \
,\nonumber\\
\nonumber\\
|z_2|^2 &=& \frac{1 }{ 2}(P_1^2 + Q_1^2 + P_2^2 + Q_2^2)+E \ ,
\end{eqnarray}
where
\begin{equation}
E\equiv Q_1 P_2 - Q_2 P_1\ .
\end{equation}
$Q_1, P_1$ and $Q_2, P_2$ are electric and magnetic charges of
a vector and axial-vector fields, respectively.

The entropy and temperature of these solutions are
\begin{equation}
T =   {1 \over 2\pi M} \quad
 {\sqrt{(M^2 - |z_1|^2) \; (M^2 - |z_2|^2)}\over \left[ \sqrt{M^2 -
|z_1|^2 } +  \sqrt{ M^2 - |z_2|^2}\right ]^2}\ .
\end{equation}

The entropy, which equals  one quarter of the area of the horizon, is

\begin{equation}
 S =  \pi \left[ \sqrt{M^2 - |z_1|^2 } +  \sqrt{ M^2 - |z_2|^2}\right
]^2
\end{equation}

Now we are in a position to discuss the extreme limit of the generic
$U(1)\times U(1)$ four-dimensional
axion-dilaton black holes,
 as well as the extreme limit to $U(1)$ black hole.

Extreme charged  $U(1)\times U(1) $ axion-dilaton black holes have a
restored
$N=1$ supersymmetry. The temperature and entropy of those solution
should be
calculated in the limit
\begin{equation}
{\cal B}_1 \rightarrow 0 \ ,\quad {\cal B}_2 \neq 0
\label{lim1}\end{equation}

or \begin{equation}
{\cal B}_2 \rightarrow 0 \ ,\quad {\cal B}_1 \neq 0
\label{lim2}\end{equation}
In both cases the extreme solution has zero temperature and non-zero
entropy,
\begin{eqnarray}
T&=&0 \ ,
\nonumber\\
S&= &\pi\; ({\cal B}_1 )^2 = \pi \;({\cal B}_2 )^2 =   \pi \;(|z_1|^2
- |z_2|^2 )
\end{eqnarray}

There are different  ways to reach the restoration of $N=2$
supersymmetry. One possibility is to consider
the next step after that in eq. (\ref{lim1}) to take
\begin{equation}
{\cal B}_2 \rightarrow 0 \ ,
\label{lim11}\end{equation}
or, if we started with eq. (\ref{lim2}),  to proceed with
\begin{equation}
{\cal B}_1 \rightarrow 0 \ .
\label{lim22}\end{equation}
In this particular limit
\begin{eqnarray}
T&=&0\ ,
\nonumber\\
S&= &0\ .
\end
{eqnarray}
However, one may first take the limit to $U(1)$ black hole by
choosing $z_1 \rightarrow z_2$, which means
\begin{equation}
{\cal B}_1 \rightarrow  {\cal B}_2 \ ,
\label{lim STW}\end{equation}
and then reach extreme by requiring
\begin{equation}
{\cal B}_1 =  {\cal B}_2 \rightarrow 0 \ .
\label{lim exSTW}\end{equation}

In this limit the temperature of an extreme $U(1)$ axion-dilaton
black hole reaches the value of the Schwarzschild one, and the
entropy vanishes, in agreement with \cite{G}, \cite{STW} .

There are many other ways to consider the limit where  both SUSY
bounds
are saturated and ${\cal B}_1 =  {\cal B}_2 =0$. The temperature is
ill defined in this limit and could take any value between zero and
${1\over8\pi M}$. The entropy, however,
in all cases approaches zero.

The analogous situation was discussed at length in ref.
\cite{US} and illustrated in figures, where we plotted both
temperature and
entropy either as the functions of central charges at fixed mass or
as the
function of mass at a given central charges.
 All figures presented there for dilaton solution apply to the more
general axion-dilaton solutions under the condition that now instead
of the
central charges the moduli of complex central charges has to be used
since only
in absence of axion are the central charges real.

After this discussion of most interesting thermodynamical properties
of axion-dilaton black holes and their relation to supersymmetry we
are going to a more technical part of the talk and present our
solutions.

Our conventions and our action are those of refs.
\cite{US} and \cite{KO}.  We
will use the complex scalar $\lambda=iz=a+ie^{-2\phi}$,
where $a$ is the axion field and $\phi$
is the dilaton field. We also have two  $U(1)$ vector
fields $A_{\mu}^{(n)}$, $n=1,2$.

We find it convenient to define the
$SL(2,R)$-duals\footnote{The spacetime duals are
${}^{\star}F^{(n)\mu\nu}=\frac{1}{2\sqrt{-g}}
\epsilon^{\mu\nu\rho\sigma}F_{\rho\sigma}$, with
$\epsilon^{0123}=\epsilon_{0123}=+i $  .} to the
fields $F_{\mu\nu}^{(n)}= \partial_{\mu} A_{\nu}^{(n)}-
\partial_{\nu}
A_{\mu}^{(n)}$\ ,
\begin{equation}
\tilde{F}^{(n)}=e^{-2\phi}\,{}^{\star}F^{(n)}-iaF^{(n)}\; ,
\end{equation}
in terms of which the action reads
\begin{equation}\label{eq:action1}
S=
\frac{1}{16\pi}
\int d^{4}x\sqrt{-g}\biggl
\{-R+\frac{1}{2}\frac{\partial_{\mu}\lambda
\partial^{\mu}\overline{\lambda}}{({\mbox{Im}} \; \lambda)^{2}}
-\sum_{n=1}^{N}F^{(n)}_{\mu\nu}{}^{\star}
\tilde{F}^{(n)\mu\nu} \biggr \}\; .
\end{equation}
In terms of the component fields, we have
\begin{eqnarray}\label{eq:action2}
S & = &
\frac{1}{16\pi}
\int d^{4}x\sqrt{-g}\biggl \{-R+2(\partial\phi)^{2}
+\frac{1}{2}e^{4\phi}(\partial a)^{2}-
\nonumber \\ &&
-e^{-2\phi}\sum_{n=1}^{N}(F^{(n)})^{2}
+ia\sum_{n=1}^{N}F^{(n)}{}^{\star}F^{(n)}\biggr \}\; .
\end{eqnarray}

The advantage of using $\tilde{F}^{(n)}$ is that the
equations of motion imply the local existence of $N$
real vector potentials $\tilde{A}^{(n)}$ such that
\begin{equation}
\tilde{F}^{(n)}=i\, d\tilde{A}^{(n)}\; .
\end{equation}
The analogous equation
$F^{(n)}=dA^{(n)}$ is not a consequence of  equations
of motion but a consequence of the Bianchi identity.
If the time-like components $A_{t}^{(n)}$
play the role of electrostatic
potentials, then the $\tilde{A}^{(n)}_{t}$  will play the role of
magnetostatic potentials. The $SL(2,R)$ duality transformations
consist in the mixing of $A^{(n)}$ with $\tilde{A}^{(n)}$ and
of equations of motion with Bianchi identities,
as in the Einstein-Maxwell case.

Here we present two different kinds of static
solutions to the equations of motion of the action
(\ref{eq:action1}), (\ref{eq:action2}): spherical
black-hole solutions and multi-black-hole
solutions, both with nontrivial axion, dilaton and
$U(1)$ fields. All the previously
known solutions of these kinds (Schwarzschild, (multi-)
Reissner-Nordstr\"{o}m, the purely electric and magnetic
dilaton black holes of refs. \cite{G},
 the electric-magnetic black holes
of refs. \cite{US} and \cite{T}, and the
axion-dilaton black holes of refs. \cite{STW}
and \cite{T}) are particular cases of them.

The static spherically symmetric black-hole solutions are \cite{KO}
\begin{eqnarray}
ds^{2}                 & = &
e^{2U}dt^{2}-e^{-2U}dr^{2}-R^{2}d\Omega^{2}\; ,
\nonumber \\
\nonumber \\
\lambda(r)             & = &
\frac{\lambda_{0}r+\overline{\lambda}_{0}\Upsilon}{r
+\Upsilon}\; ,
\nonumber \\
\nonumber \\
A_{t}^{(n)}(r)         & = &
e^{\phi_{0}}R^{-2}[\Gamma^{(n)}(r+\Upsilon)+c.c]\; ,
\nonumber \\
\nonumber \\
\tilde{A}_{t}^{(n)}(r) & = &
-e^{\phi_{0}}R^{-2}[\Gamma^{(n)}(\lambda_{0}r+
\overline{\lambda}_{0} \Upsilon)+c.c]\; ,
\end{eqnarray}
where
\begin{eqnarray}
e^{2U}(r)  & = & R^{-2}(r-r_{+})(r-r_{-})\;,  \qquad
r_{\pm}    = M\pm r_{0}\; ,
\nonumber \\
R^{2}(r)   & = &  r^{2}-|\Upsilon|^{2}\; ,
\hspace{2,8 cm}r_{0}^{2}=M^{2}+|\Upsilon|^{2}
-4\sum_{n=1}^{N} |\Gamma^{(n)}|^{2}\; .
\end{eqnarray}

We define the parameters of our solutions
in terms of the asymptotic behavior ($r\rightarrow
\infty$) of the different complex fields
\begin{eqnarray}
g_{tt} & \sim  & 1-\frac{2M}{r}\; ,
\hspace{3cm}
\lambda \sim \lambda_{0}-ie^{-2\phi_{0}}
\frac{2\Upsilon}{r}\; , \nonumber \\
F_{tr}  & \sim
&\frac{e^{+\phi_{0}}Q}{r^{2}}\; . \hspace{3.3cm}
{}^{\star}F_{tr}  \sim \frac{i
e^{+\phi_{0}}P}{r^{2}}\; .
\end{eqnarray}

The real axion ($\Delta$), dilaton ($\Sigma$), electric
($Q$) and magnetic ($P$) charges, and the
asymptotic values of the
axion ($a_{0}$) and dilaton ($\phi_{0}$) are
\begin{equation}
\Upsilon  = \Sigma-i\Delta,
\hspace{1cm}
\Gamma   = \frac{1}{2}(Q+iP),
\hspace{1cm}
\lambda_{0}  =  a_{0}+ie^{-2\phi_{0}}.
\end{equation}

In every black hole in our solutions the charge of
the complex scalar  $\Upsilon$ is related to the electromagnetic
charges by
\begin{equation}
\Upsilon=- \frac{2
}{M}\, (\overline{\Gamma}_{n})^{2}\; .
\end{equation}

The singularity is hidden under a horizon if $r_{0}^{2}>
0$, and it is hidden or coincides with it (but still is
invisible for external observers) if $r_{0}=0$.
The  conditions
$r_{0}^{2} \geq 0$
 and $M\geq |\Upsilon|$ can be related to supersymmetry bounds
\cite{US}, \cite{T}, \cite{FUTURE}. All solutions given
above
have the entropy
\begin{equation}
S= \pi (r_{+}^2 - |\Upsilon|^{2}) \ .
\end{equation}
When all  supersymmetric bounds are saturated, i.e.
$r_{+}=M=|\Upsilon|$,
the objects described by this solution have
 zero area of the horizon and vanishing entropy.  In this sence, such
black holes ({\it holons})  behave as elementary particles.

Our second
kind of solutions describe  axion-dilaton extreme multi-black-hole
solutions \cite{KO}.
 The fields are
\begin{eqnarray}
ds^{2}           & = & e^{2U}dt^{2}-e^{-2U}d\vec{x}^{2}\;
, \hspace{1cm}
                                e^{-2U}(\vec{x})=2 \;{\mbox{Im}}\;
                             ({\cal H}_{1}(\vec{x})\;
                                 \overline{{\cal
                                H}}_{2}(\vec{x}))\; ,
                                    \nonumber \\
\nonumber \\
\lambda(\vec{x}) & = & \frac{{\cal H}_{1}(\vec{x})}{{\cal
                         H}_{2}(\vec{x})}\; ,
                        \nonumber \\
\nonumber \\
A_{t}^{(n)}(\vec{x})       & = & e^{2U}(k^{(n)}{\cal
                                H}_{2}(\vec{x})+c.c)\; ,
                        \nonumber \\
\nonumber \\
\tilde{A}^{(n)}_{t}(\vec{x}) & = & -e^{2U}(k^{(n)}{\cal
                                 H}_{1}(\vec{x})+c.c)\; ,
\end{eqnarray}
where ${\cal H}_{1}(\vec{x}), {\cal H}_{2}(\vec{x})$
are two complex harmonic functions
\begin{eqnarray}
{\cal H}_{1}(\vec{x}) & = & \frac{e^{\phi_{0}}}{\sqrt{2}}
                            \{\lambda_{0}+\sum_{i=1}^{I}
                            \frac{\lambda_{0}M_{i}+
                            \overline{\lambda}_{0}
                            \Upsilon_{i}}{|\vec{x}
                            -\vec{x}_{i}|}\}\; ,
                                      \nonumber \\
\nonumber \\
{\cal H}_{2}(\vec{x}) & = & \frac{e^{\phi_{0}}}{\sqrt{2}}
                            \{1+\sum_{i=1}^{I}\frac{M_{i}+
                            \Upsilon_{i}}{|\vec{x}
                            -\vec{x}_{i}|}\}\; .
\end{eqnarray}

The horizon of the $i$-th (extreme) black hole is at
$\vec{x}_{i}$ (in these
isotropic coordinates the horizons look like single
points), and has mass $M_{i}$, electromagnetic
charge $\Gamma_{i}$, etc. , as can be seen by using the
definitions in the Appendix
in the limit $|\vec{x}-\vec{x}_{i}|\rightarrow\infty$.
Charges without
label are total charges. The constants $k^{(n)}$ are
\begin{equation}
k^{(n)}=
-\sqrt{2}
\biggl (\frac{\Gamma^{(n)}M+\overline{\Gamma^{(n)}
\Upsilon}}{M^{2}-|\Upsilon|^{2}}\biggr )\; .
\end{equation}
The consistency of the solution requires for every $i$
\begin{equation}
k^{(n)}_{i}=k^{(n)}\; ,
\hspace{1cm}
Arg(\Upsilon_{i})=Arg(\Upsilon)\; .
\end{equation}
Finally, for each $i$ and also for the total charges, the
supersymmetric Bogomolny bound is saturated:
\begin{equation}
M^{2}+|\Upsilon|^{2}-
4|\Gamma^{(n)}|^{2}=0\; .
\end{equation}

For a single $U(1)$ vector field
 the extreme solution simplifies to ($k\equiv k^{1}$, $\Gamma \equiv
\Gamma^{1}$)
\begin{equation}
M^{2}=|\Upsilon|^{2}\ , \quad k
 =  -\frac{ \sqrt{2}  \;\Gamma}{ M}
\; .\end{equation}
As a consequence
of all the identities obeyed by the charges, it is
possible to derive the following expression of equilibrium
of forces between two extreme black holes \cite{FUTURE}:
\begin{equation}\label{eq:equiforce}
M_{1}M_{2}+\Sigma_{1}\Sigma_{2}+\Delta_{1}\Delta_{2}=
Q_{1}Q_{2}+
P_{1}P_{2}\; .
\end{equation}

 The possibility that black holes may quantum mechanically split into
other
black holes was proposed in  \cite{US}. Splitting of black
holes is
closely related to the possibility of splitting of the  universe into
many baby
universes. A particularly relevant example is splitting of one
Robinson-Bertotti universe into many RB universes, as discussed by
Brill
\cite{Brill}. For a recent discussion of splitting of dilaton black
holes with
massive dilaton fields see  \cite{Hor}.   A specific example
of the splitting of the extreme
electric-magnetic black hole into a purely magnetic and a purely
electric one was considered in \cite{KOP}.
 Such bifurcation  is forbidden classically but  could in principle
occur in a
quantum-mechanical process and may be enforced by quantum-mechanical
instability of the zero temperature state with  non-integer value of
$e^{-S}$ .  As the result, the black holes with $S\neq 0$  may split
into holons with
$S = 0$ and zero area of the horizon. Will the  black holes
continue
 splitting
to the smallest values of charges? These and
many other questions can be asked in connection with stringy
 black holes and their supersymmetric properties.

\end{document}